# All-optical quantum signal demultiplexer

Yin-Hai Li,[1,2,+] Wen-Tan Fang,[3,+] Zhi-Yuan Zhou,[1,2,*] Shi-Long Liu,[1,2] Shi-Kai Liu,[1,2] Zhao-Huai Xu,[1,2] Chen Yang,[1,2] Yan Li,[1,2] Li-Xin Xu,[3] Guang-Can Guo,[1,2] and Bao-Sen Shi[1,2,*]

[1]*CAS Key Laboratory of Quantum Information, USTC, Hefei, Anhui 230026, China*

[2]*Synergetic Innovation Center of Quantum Information & Quantum Physics,*
*University of Science and Technology of China, Hefei, Anhui 230026, China*

[3]*Department of Optics and Optical Engineering, University of Science and Technology of China,*
*Hefei, Anhui 230026, China*

*\*zyzhouphy@ustc.edu.cn; drshi@ustc.edu.cn*

Dense wavelength division multiplexing (DWDM) is one of the most successful methods for enhancing data transmission rates in both classical and quantum communication networks. Although signal multiplexing and demultiplexing are equally important, traditional multiplexing and demultiplexing methods are based on passive devices such as arrayed waveguides and fiber Bragg cascade filters, which, although widely used in commercial devices, lack any active tuning ability. In this work, we propose a signal demultiplexing method based on sum frequency generation (SFG) with two significant features: first, any signal from the common communication channel can be demultiplexed to a single user by switching the pump wavelength; second, a cheap high-performance detector can be used for signal detection. These two features were demonstrated by demultiplexing multi-channel energy-time entanglement generated by a micro-cavity silicon chip. High interference visibilities over three channels after demultiplexing showed that entanglement was preserved and verified the high performance of the demultiplexer, which will find wide application in high-capacity quantum communication networks.

## I. INTRODUCTION

Rapid developments in both classical and quantum optical communications require new optical devices and distinct methods for enhancing the communication capacity through a single communication channel [1,2]. Various signal multiplexing methods have been constructed to enhance the transmission capacity in optical communications that make use of the different degrees of freedom of light, such as time, frequency, polarization, amplitude, phase and orbital angular momentum. Some typical signal multiplexing methods include time-division multiplexing, wavelength-division multiplexing (WDM), polarization-division multiplexing (PDM), quadrature amplitude modulation and space-division multiplexing [3]. These methods are not only useful for enhancing the data transmission rate in classical optical communications; they are also the key to enhancing information transmission capacity in quantum communications. From among these signal multiplexing methods, WDM is the most widely used in commercial optical communication networks. In the area of quantum communications, dense wavelength division multiplexing (DWDM) is widely used for quantum key distribution [4,5] and entanglement distribution [6–10].

Signal multiplexing and demultiplexing are two equally important parts of the DWDM technique. Traditional multiplexing and demultiplexing methods are based on passive devices such as arrayed waveguide grating filters and fiber Bragg grating cascade filters. Such passive devices, although widely used in commercial devices, lack any active tuning ability. Recently, optical signal processing methods based on nonlinear processes such as sum-frequency generation (SFG) are being widely used in both classical and quantum regimes [11–18]. SFG preserves the quantum coherence while translating the signal from one wavelength to another. Such processes can be used for up-conversion detection and to build a frequency interface with quantum memory [12,13] or an underwater communication channel [19]. Such processes can also be used for temporal mode sorting in high dimensional systems for high-speed quantum communications [18]. Because of the phase matching condition in SFG, this process is a natural bandpass filter, and the central wavelength and bandwidth of the filter can be freely engineered by adjust the parameters of the nonlinear crystal and pump laser wavelength. In this work, we propose an all-optical tunable bandpass filter that utilizes a totally different signal demultiplexing method based on SFG. The SFG-based demultiplexer has two significant features: first, any signal from the common communication channel can be demultiplexed to a single user by switching the pump wavelength; second, a higher performance and cheaper detector can be used for signal detection. The SFG signal can also be interfaced with other quantum systems such as quantum memory or an underwater communication channel.

To verify these two distinct features of the SFG-based demultiplexer, multi-channel energy-time entanglement states are generated based on a spontaneous four-wave mixing (SFWM) process in a silicon based micro-cavity. The free spectral range (FSR) of the micro-cavity is designed to match the International Telecommunication Union's (ITU) standard for fiber communication channels. This strategy permits distributing energy-time entanglement simultaneously over multi-channels. Cavity-enhanced SFWM is used to generate narrow bandwidth photon pairs with enhanced spectral brightness, which is rather different compared with single pass configurations [10]. In the cavity-enhanced case, the bandwidth of the photon pair is well within the SFG bandwidth, and therefore all the photons can be up-converted with equal efficiency, while for single-pass configurations, only part of the photon lies within the effective SFG bandwidth. Three pairs of energy-time entanglement states with different frequencies are up-converted and demultiplexed to a visible wavelength in a single resonant cavity configuration using different pump wavelengths. The quantum conversion efficiency can reach 38 %. High visibilities between up-converted photons and the paired frequency-unchanged photons indicate the preservation of photonic entanglement and verifies the high performance of our proposed quantum signal demultiplexer. The present work is very promising for high-speed quantum communications.

## II. PRINCIPLE OF THE ALL-OPTICAL TUNABLE QUANTUM DEMULTIPLEXER

Figure 1(a) illustrates the principle of the proposed active quantum signal demultiplexer. S1-S3 represent multi-frequency energy-time entangled photon pairs with 100-GHz frequency spacings that transmit in a common communication channel. Frequency up-conversion based on SFG is used to up-convert the signal photon to a visible wavelength and demultiplex it from the common communication channel. Different signal photon frequency modes can be addressed simply by switching the pump wavelength to achieve the quasi-phase matching condition; signal frequency modes whose phases do not match the pump wavelength are not up-converted. Preservation of entanglement and the performance of the quantum signal demultiplexer can be verified by measuring the correlation between the up-converted visible signal photons and their paired idler photons. The up-converted signal photons can be used for further processing, including coupling to a quantum memory and interfacing with an underwater communication channel, or can be directly detected with cheap high-performance silicon single photon detectors.

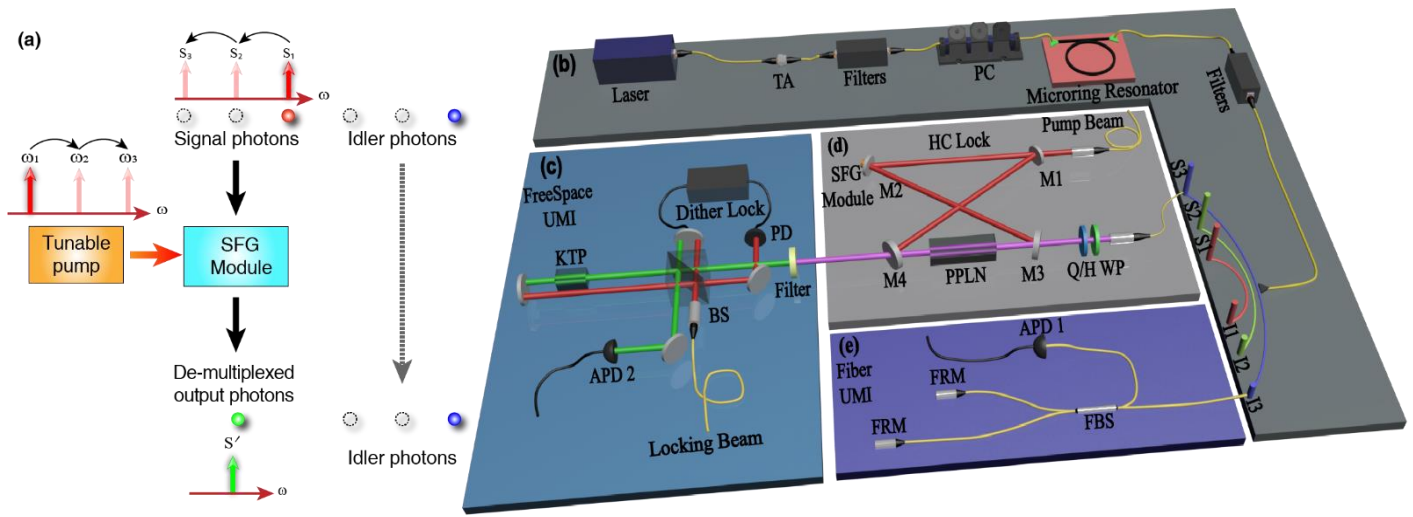

Fig. 1. Basic principle and experimental setup of the quantum signal demultiplexer. (a) Principle behind using SFG for the active demultiplexing of multiplexed entanglement states. (b) Laser; TA: tunable attenuator; Filters: 100-GHz WDM cascade filters; PC: polarization controller; Microring Resonator: multiplexed energy-time entangled photon source. (c) Free-Space UMI; BS: beam splitter; KTP: potassium titanyl phosphate; APD2: Si avalanched single photon detector; Locking Beam: small part of the 795-nm pump laser, aimed at locking the free space UMI; PD:    (d) Cavity-enhanced SFG module; M1-M4, cavity mirrors; Q(H)WP: quarter (half) wave plate; PPLN: periodically-poled lithium niobate; Pump Beam: 795-nm laser for pumping the cavity; HC lock: Hänsch-Couillaud lock. (e) Fiber UMI; APD1: ID220 free-running (In,Ga)As avalanched single photon detector; FRM: Faraday rotator mirror; FBS: fiber beam splitter. The frequency conversion is performed in a ring cavity, which is pumped at 795 nm.

### III. EXPERIMENTAL SETUP

Figure 1(b)–(e) illustrate the experimental setup. As shown in Fig. 1(b), a silicon waveguide microring resonator was employed for multi-frequency energy-time entanglement generation based on SFWM. The ring resonator had a high quality (Q) factor of 430,000 and a bandwidth of 490 MHz. The FSR of the cavity was 200 GHz. Figure 2(a) shows a typical transmission profile for the microring. Pump light from a wavelength-tunable continuous-wave (CW) diode laser (DL pro, Toptica Photonics AG Germany) was first filtered by 100-GHz cascade DWDM filters to remove the broadband background fluorescence. The polarization of the pumped light was aligned to the TE mode of the silicon waveguide using a polarization controller (PC). We used a temperature controller to control the resonant frequency of the ring cavity to tune the cavity resonant frequency to the pump frequency. It has been shown that the resonant frequencies are shifted by ~10 GHz/K due to thermo-optic effects in the silicon substrate [20]. With the cleaned pump laser injected into the chip, photon pairs were generated through the SFWM and separated by a 32-channel DWDM device. The corresponding channels at equal intervals from the pump frequency were correlated [21]. In our experiment, we fixed the pump wavelength at 1550.12 nm (ITU channel 34) and chose three pairs of channels with low noise and high brightness photon pairs in the ITU grid (see Fig. 2(a)). We relabelled the signal and idle photons S1-S3 and I1-I3 around the central wavelength of the ITU grid; see Table II in Appendix A.

The microring was coupled to a bus waveguide located on one side of the ring. Both the ring and bus waveguides were etched on SOI substrate, with the same transverse dimensions of 220 nm (height) ×450 nm (width). The pump laser was coupled to the waveguide through input-coupling gratings, with total insertion losses of about 10 dB from the input to output ports. Three paired channels were chosen in the experiment and labelled as in Fig. 1(b). After separating with DWDM filters, the signal photon was sent to the SFG module, which contained a single resonant cavity for 795 nm. The infrared and up-converted visible single photons passed through the cavity, which contained a type-0 phase-matched periodically-poled lithium niobate (PPLN) crystal designed for a frequency conversion of 795 nm + 1560 nm → 525 nm. The strong pump laser was a Ti:sapphire laser (MBR110, Coherent, Santa Clara, California, United States). The cavity was actively locking using the Hänsch-Couillaud technique [22]. After up-conversion, the visible photons were injected into a free space unbalanced Michelson interferometer (UMI) with a 1.6 ns time difference between the two unbalanced arms, which had the same optical path length difference as the fiber UMI for the idler photons. The free space UMI was locked to the 795-nm pump beam using dither locking with a 5-kHz dither frequency. The phase of the free space UMI was tuned by controlling the temperature of the KTP crystal. For details of the free space UMI, please refer to the Appendix D. At the output of the free space UMI, these 525-nm photons were coupled into a single-mode fiber with 60 % coupling efficiency. The photons output from the free space UMI and fiber UMI were detected by two single photon detectors (APD2: Si single photon detector, 50 % detection efficiency; APD1: ID220, InGaAs single photon detector, 20 % detection efficiency, 5 us dead time). Output electronic signals were processed and sent to the coincidence device (TimeHarp 260, PicoQuant, Berlin, Germany, 0.8-ns detection window) for measurement.

### IV. CHARACTERIZING THE PERFORMANCE OF THE QUANTUM SIGNAL DEMULTIPLEXER

To characterize the performance of the demultiplexer, it is most important to verify the preservation of quantum entanglement after up-conversion. Before moving to this final target, the parameters of the microring-based photon source and the up-conversion SFG module need to be characterized. The accidental coincidence ratio (CAR) against pump power was measured first, and the results are shown in Fig. 2(b). The CAR has the same shape as reported for a single silicon nanowire [10], but the pump power to reach the maximum CAR is greatly reduced, which arises from the cavity-enhanced effects. The reason for the increase in CAR at low pump power and the decrease in CAR at higher pump power is addressed in Ref [23]. To characterize the performance of the SFG module, the quantum efficiency of the SFG module was measured using a narrow bandwidth diode laser (DL Pro, Toptica Photonics AG; bandwidth 1 MHz; tuning range of 1520 nm to 1590 nm). We fixed the pump power of the diode laser at 1 mW, to measure the up-converted SFG power as a function of the 795-nm pump power. Figure 2(c) shows the results; the power conversion efficiency is defined as $\eta_{power} = p_{525}/p_{1560}$ and the corresponding quantum conversion efficiencies can be expressed as $\eta_{quantum} = \eta_{power}\lambda_{525}/\lambda_{1560}$. The maximum quantum efficiency obtained was 38 % for a pump power of 550 mW.

We measured the CAR of the up-converted photon pair as a function of the pump power injected into the silicon chip for channels I2–S'2 ( represent the up-converted photons corresponding to signal photons ) with the 795-nm pump power fixed at 400 mW. The coincidence time window was still 0.8 ns. The CAR changed as shown in Fig. 2(d), the difference was that the pump power required to reach maximum CAR shifted from 150 μW to 400 μW. Because of losses in the experiment and the relative low quantum conversion efficiency, the dark count level for APD2 (~1000/s) was dominant in the measurement (the number of up-converted photons was approximately 2000/s).

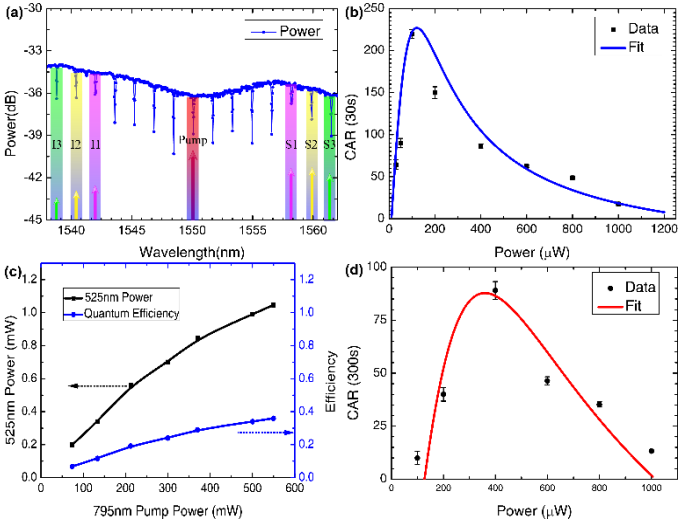

Fig. 2 Characterizing the basic parameters of the microcavity-based multi-frequency energy-time entangled sources. (a) Transmission spectrum of the microcavity measured with an optical spectrum analyzer. (b) CAR as a function of the pump power for channel pairs I2–S2. (c) Quantum efficiency and 525-nm SFG power as functions of the 795-nm pump power. (d) CAR between idler photon I2 and up-converted signal photon S'2.

Next, the performance of the demultiplexer was verified by showing its ability to preserve the energy-time entanglement and the ability to freely select the signal wavelength by switching the pump wavelength. The quality of the multi-frequency energy-time entanglement before frequency transduction was measured first. We then measured the energy-time entanglement between the up-converted photons and the corresponding idler photons.

These results are grouped and compared in Fig. 3. By post-selection of the central peak of the coincidences using a proper time window, the energy-time entangled state can be expressed as ( $S$ and $L$ represent photons that pass the short and long arms of the UMIs respectively):

$$|\Phi\rangle_{in} = \frac{1}{\sqrt{2}}(|SS\rangle + |LL\rangle), \quad (1)$$

Figure 3 shows the two-photon interference fringes before and after up-conversion for signal and idler pairs and Table I summarizes the corresponding raw and net visibilities.

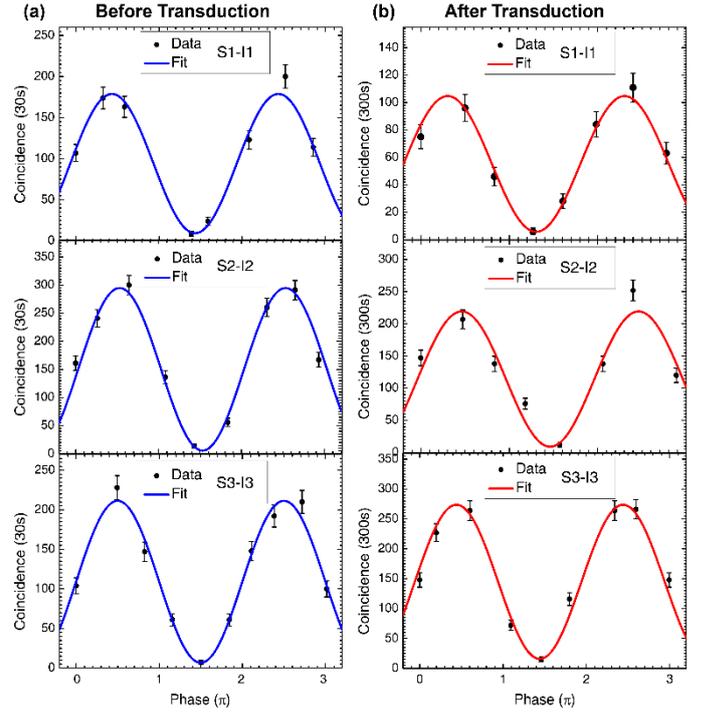

Fig. 3. Two photon interference fringes before and after quantum frequency conversion. The coincidence time was 30 s before quantum frequency conversion (QFC) and 300 s after QFC. Error bars were added assuming that the photon counting process follows a Poisson distribution.

Table I. Raw and net visibilities before and after QFC

| Channel pairs | Before QFC | | After QFC | |
|---|---|---|---|---|
| | Raw visibility | Net visibility | Raw visibility | Net visibility |
| **S1-I1** | (89.98 ± 3.82) % | ( 93.98 ± 2.95) % | (89.69 ± 5.07) % | (96.69 ± 2.84)% |
| **S2-I2** | (96.22 ± 1.77) % | ( 98.86 ± 0.94) % | (91.98 ± 3.07) % | (95.32 ± 2.33) % |
| **S3-I3** | (93.82 ± 2.72) % | ( 97.39 ± 1.73) % | (89.02 ± 3.23) % | (91.56 ± 2.84) % |

It is clear from Table I that all the raw and net visibilities were approximately 90 %; the up-conversion process has little influence on the interference visibility. High visibilities much greater than 71 % also indicate the possibility of violating Bell's inequality, which shows the preservation of quantum entanglement in a tunable demultiplexer. Before frequency conversion, the measured cumulative coincidence time was 30 s, and after frequency conversion it was 300 s. The error bars were obtained assuming that the photon counting process follows a Poisson distribution. The net visibilities were obtained by subtracting the dark coincidences in the coincidence measurement; the dark coincidences were obtained by recording coincidence counts far away from the central time window. In all cases, the pump power just before injection into the waveguide was 400 μW. A second single photon detector (APD2) and fiber UMI with the same parameters as those used for the idler photons were used to measure the interference for the source. To demultiplex the different frequency modes of the energy-time entanglement state, the SFG pump wavelength was switched to a value that met the phase matching condition.

Finally, we would like to give an estimation of the overall detection efficiencies for idler photons and up-converted signal photons. The insertion loss of the silicon waveguide was 5.00 dB for both photons and the filtering loss of the cascade DWDM filters was about 2.00 dB for both photons. The total losses for the signal photons (from the input port of the SFG module) were 8.59 dB, which includes transmission loss (0.8 dB), up-conversion loss (5.38 dB), filtering loss (0.20 dB) and fiber coupling loss (2.21 dB). Taking into account the detection efficiency of the InGaAs (20 %, 6.99 dB) and the Si (50 %, 3.00 dB) single photon detector, the overall detection efficiencies for the idler and signal photons were 13.99 dB and 15.59 dB respectively.

## V. CONCLUSION AND DISCUSSION

This work provides a roadmap for an all-optical quantum signal demultiplexers for multi-frequency energy-time entanglement states generated by a silicon microcavity. The proposed quantum signal demultiplexer has two distinct features: any signal encoded in the frequency modes can be dropped out from a common communication channel by switching the pump laser frequency in the SFG process; the demultiplexed signal can be directly detected with a cheap high-performance single photon detector and interfaced with quantum memories or an underwater communication channel. High visibilities of two-photon interference over three channels after demultiplexing clearly proved that entanglement was fully preserved in the SFG process. The scheme provides a new optical device that is fully compatible with the C-band DWDM technique, and may be of great importance for future high-capacity quantum networks. Further improvements to the present system will include increasing the conversion efficiency in the up-conversion process, optimizing the efficient SFG bandwidth by adjusting the parameters of the crystal, reducing the insertion loss of the waveguide by special grating design and encapsulating the device for a minimized quantum system. Once these issues have been efficiently solved, the quantum signal demultiplexer will find wide application in high-speed high-capacity quantum communications.

## ACKNOWLEDGMENTS

This work was supported by the National Natural Science Foundation of China (NSFC) (11174271, 11604322, 61275115, 61435011, 61525504, 61605194), the China Postdoctoral Science Foundation (2016M590570, 2017M622003), the National Key Research and Development Program of China (2016YFA0302600), and Fundamental Research Funds for the Central Universities.

Y.-H. L. and W.-T. F. contributed equally to this work.

## APPENDIX A: DEFINITION OF THE WAVELENGTHS OF THE STANDARD ITU GRIDS

The corresponding wavelength for the correlated signal and idle photons are defined in Table II. The bolded channels are used for entanglement measurements of the multichannel performance in the experiments. The pump wavelength is located at the center of channel C34.

Table II. Definition of the wavelengths of the standard ITU grids for the signal and idler photons

| Name | DWDM channel | Wavelength ( nm ) |
| --- | --- | --- |
| **Signal 3 - Idler 3** | C20 - C48 | 1561.42 – 1538.98 |
| Signal 2 - Idler 2 | C22 – C46 | 1559.79 – 1540.56 |
| **Signal 1 – Idler1** | C24 – C44 | 1558.17 – 1542.14 |
| **Pump** | C34 | 1550.12 |

## APPENDIX B: MORE DATA ON THE ENERGY-TIME ENTANGLED SOURCE

We measured the FWHM of the silicon micro-cavity using a wavelength tunable diode laser with 10MHz bandwidth. We measure the transmission power versus frequency shift of the pump laser. The results is shown in Fig. 4 (a), where we can see the FWHM of micro-cavity is 490MHz. Fig. 4 (b) shows single count rates as a function of the pump power for channels s2 and i2. The single count rates are increasing squared with the increasing of the pump power. The difference in the count rates between channels s2 and i2 are caused by Raman-scattering photons come from pigtails of filters.

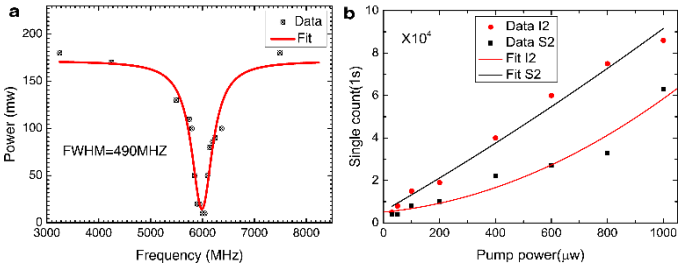

Fig. 4 (a) Transmission power as a function of frequency shift of the pump laser; (b) single photon count rates as a function of pump power for signal and idler channels S2 and I2.

## APPENDIX C: SFG CAVITY DESIGN

The bow-tie ring cavity is designed for a single resonance at 795nm, and the total cavity length is 547mm. The input coupling mirror M1 has transmittance of 3% at 795nm. Mirror M2 highly reflectively coated at 795nm (R>99.9%), and a piezoelectric element (PZT) is attached to it to scan and lock the cavity. The two concave mirrors, M3 and M4, both have curvatures of 80mm; M3 has a high transmittance coating for 1560nm (T>99%) and is highly reflectively coated for 795nm (R>99.9%), while M4 has a high transmittance coating for 525nm (T>98%) and is highly reflectively coated for 1560nm and 795nm (R>99.9%). The fundamental cavity mode has a beam waist of 60 μm at the mid-point of mirrors M3 and M4. The PPLN crystal we used has dimensions of 1mm×0.5mm×50mm, both end faces are anti-reflected coated for 795nm, 525 nm and 1560nm. The crystal has a poling period of 7.3 μm. The measured phase matching temperature is 29.5°C.

## APPENDIX D: HOW THE UMI PHASE IS TUNED IN THE EXPERIMENT

In the fiber UMI, the thermal coefficient of fiber at 1550 nm is $\frac{dn}{dT} = 0.811 \times 10^{-5} / °C$, the fiber length difference of the UMI is (163.48 mm) $L_d = c\Delta t / 2n$ for 1.6 ns delay. The temperature for one tuning period $\Delta T = \lambda / (2 L_d \frac{dn}{dT})$ is 0.585K. In the free-space UMI, the thermal coefficient of KTP at 525 nm is $\frac{dn_z}{dT} = 1.6 \times 10^{-5} / °C$, the fiber length difference of the UMI is (163.48 mm) $L_d = c\Delta t / 2n$ for 1.6 ns delay. The temperature for one tuning period $\Delta T = \lambda / (2 L_d \frac{dn}{dT})$ is 1.16K. In the experiments, the temperature tuning periods and the phase of the both fiber and free-space UMI are measured and calibrated using a stable narrow bandwidth laser source. The phase of the UMIs can keep unchanging for hours because of seriously thermal and acoustic isolation from the environment.